\documentclass{WileyMSP-template}

\begin{document}

\theoremstyle{plain}          
\newtheorem{theorem}{Theorem}[section]
\newtheorem{lemma}{Lemma}[section]
\newtheorem{definition}{Definition}[section]
\newtheorem{proposition}{Proposition}[section]
\newtheorem{conjecture}{Conjecture}[section]
\newtheorem{question}{Question}[section]
\newtheorem{corollary}{Corollary}[section]

\theoremstyle{definition} 
\newtheorem{example}{Example}[section]

\newcommand{\ket}[1]{{|} #1\rangle}
\newcommand{\bra}[1]{\langle #1{|}}

\pagestyle{fancy}
\rhead{\includegraphics[width=2.5cm]{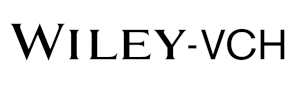}}

\title{Connecting Quantum Contextuality and Nonlocality}

\maketitle


\author{Jianqi Sheng$^{1, 2,*}$}
\author{Dongkai Zhang$^{1, \dagger}$}
\author{Lixiang Chen$^{1, \ddagger}$}


\begin{affiliations}
$^{1}$Department of Physics, Xiamen University, Xiamen 361005, China\\
$^{2}$Department of Physics, City University of Hong Kong, Kowloon, Hong Kong\\
*sheng.jq@city.edu.hk\\
$\dagger$ zhangdk@hqu.edu.cn\\
$\ddagger$ chenlx@xmu.edu.cn\\
\end{affiliations}


\keywords{quantum correlations, nonlocality, contextuality, sheaf-theory, graph-theory}

\justifying
\begin{abstract}

Quantum theory departs from classical physics in its treatment of correlations, most prominently through the phenomena of contextuality and nonlocality. 
Once regarded primarily as foundational curiosities, these effects are now understood as key operational resources for quantum computation, communication, and simulation. 
Although traditionally investigated in distinct settings, recent theoretical and experimental advances have revealed deep conceptual, mathematical, and operational connections between them. 
This review presents a unified perspective on these developments based on sheaf-theoretic and graph-theoretic frameworks, which provide theory-independent characterizations of statistical correlations. 
These approaches clarify the structural relationship between contextuality and nonlocality, facilitate the formulation of experimentally testable inequalities, and guide implementations in realistic physical platforms, with particular emphasis on photonic systems. 
By bridging abstract theoretical structures and concrete experimental realizations, this review sheds light on the nonclassical foundations of quantum correlations and their emerging role in quantum technologies.

\end{abstract}


\section{Introduction}

Quantum theory departs sharply from classical intuitions about physical reality, most notably in its treatment of correlations. Among its most profound departures are nonlocality and contextuality, which reveal fundamental limitations of classical descriptions based on locality and noncontextual realism. From a foundational perspective, these phenomena expose the impossibility of reproducing quantum correlations within broad classes of hidden-variable models. At the same time, they have proven to be far more than conceptual curiosities. Since Bell’s seminal work \cite{bell1964einstein}, nonlocal correlations have come to be understood as operational resources that underpin quantum information processing, with landmark achievements such as Shor’s polynomial-time factoring algorithm~\cite{shor1994algorithms}, quantum cryptography~\cite{bennett1984update}, and quantum teleportation~\cite{bennett1993teleporting}. More recently, growing experimental control has enabled systematic tests of quantum contextuality~\cite{bartosik2009experimental,kirchmair2009state,guhne2010compatibility,amselem2009state,lapkiewicz2011experimental,moussa2010testing,vanDam2019}, for which one might hope similar applications can be found, motivating renewed interest in its potential operational significance. 

Bell nonlocality manifests in spatially separated systems whose joint statistics violate Bell inequalities. 
Bell’s theorem~\cite{bell1964einstein} establishes that no model satisfying local causality can reproduce the full range of correlations predicted by quantum theory, while still respecting the no-signaling principle required by relativistic causality. In a standard Bell scenario, independent measurement choices are performed at distant locations, and the resulting correlations defy any decomposition into local hidden variables \cite{brunner2014bell}. Contextuality, by contrast, can arise at the level of a single quantum system. The Kochen-Specker (KS) theorem~\cite{kochen1967specker} demonstrates that, for systems of dimension $d \geq 3$, it is impossible to assign predetermined, context-independent values to all observables in a way that is consistent with quantum predictions. This shows that all noncontextual hidden-variable models cannot reproduce the statistics of compatible measurements across different contexts. Motivated by the idealized assumptions underlying this framework, Spekkens introduced a generalized notion of contextuality applicable to arbitrary operational theories~\cite{spekkens2005contextuality}, in which contextuality is identified with the impossibility of a context-independent ontological description of preparations, transformations, and measurements. 
Nonlocality and contextuality are differ in their operational requirements. Bell nonlocality requires spatially separated parties and that measurement statistics respect the ``no-signaling'' condition: the choice of measurement on one side cannot influence the marginal statistics on the other, thereby precluding superluminal communication and ensuring consistency with relativistic causality. In contrast, contextuality can arise in single systems and is governed by the compatibility structure of measurements together with the ``no-disturbance'' condition, which demands that the marginal statistics of a measurement remain invariant under the choice of jointly performed compatible observables. 

Although operationally distinct, nonlocality and contextuality are conceptually closely related. 
Since the 1970s, their conceptual relationship has been extensively investigated~\cite{stairs1978quantum,heywood1983nonlocality,stairs1983quantum,redhead1987incompleteness,brown1990nonlocality,elby1990nonlocality,elby1990physical,mermin1990simple,krips1992metaphysics,clifton1993getting,kernaghan1995kochen}. 
Early results in this direction include the KS-with-locality theorems~\cite{heywood1983nonlocality,stairs1983quantum,redhead1987incompleteness,brown1990nonlocality}, which later inspired the formulation of the free will theorem~\cite{conway2006free,kochen2022free}. 
A second wave of results was motivated by the Greenberger-Horne-Zeilinger (GHZ) proof of Bell's theorem~\cite{greenberger1989bell}. 
Mermin showed that the GHZ argument can be reformulated both as a tripartite Bell inequality and as a state-independent proof of the KS theorem~\cite{mermin1990simple,mermin1990extreme,mermin1993hidden}. 
Subsequently, Hardy's proof of Bell nonlocality was recognized as a state-dependent KS-type argument~\cite{hardy1992quantum,hardy1993nonlocality,cabello1996bell}. 
Related ideas culminated in GHZ-like constructions for two parties sharing qubits~\cite{cabello2001bell}, which can be traced back to the Peres-Mermin KS proof~\cite{mermin1990simple,peres1990incompatible} and can be translated into bipartite Bell inequalities~\cite{cabello2001all}. 
These developments gave rise to a broad family of so-called ``all-versus-nothing'' proofs~\cite{cabello2001all}, as well as to alternative formulations in terms of nonlocal games~\cite{cleve2004consequences} and quantum pseudotelepathy~\cite{renner2004proceedings,brassard2005quantum}. 
Building on these foundational insights, recent studies have further explored the connections between contextuality and nonlocality. 
Plavala~\cite{plavala2024contextuality} demonstrated that bipartite entanglement necessitates underlying preparation and measurement contextuality, establishing that contextuality is a prerequisite for observing nonlocal correlations in distributed systems. 
Complementary theoretical and experimental studies have converted contextuality into nonlocality, highlighting its practical role as a resource~\cite{cabello2021converting,sheng2025orbital}. 
Other investigations have constructed inequalities that simultaneously witness coherence, nonlocality, and contextuality, emphasizing the interdependent structure of these quantum features~\cite{wagner2024inequalities}, and identified quantum states capable of simultaneously violating Bell and noncontextuality inequalities, revealing trade-offs beyond conventional monogamy constraints~\cite{kitajima2025states}. 
Further studies have explored the interplay between contextuality, nonlocality, and Gleason’s theorem in multiqubit systems, providing insights into models of multiqubit quantum computation~\cite{wright2023contextuality}. 
These developments underscore the need for formal frameworks that capture the structural features underlying both contextuality and nonlocality, moving beyond case-by-case analyses toward a unified theoretical understanding~\cite{abramsky2011sheaf,fritz2013local,cabello2014graph,acin2015combinatorial}.

From a more general perspective, several systematic frameworks have been developed to unify contextuality and nonlocality at a structural level. 
The sheaf-theoretic approach~\cite{abramsky2011sheaf} characterizes both phenomena in terms of obstructions to global assignments of outcomes, providing a representation-independent description of quantum correlations. 
Complementarily, graph-theoretic methods~\cite{cabello2014graph} encode exclusivity relations between measurement events in combinatorial structures, yielding operationally meaningful inequalities that are particularly well suited for experimental implementation. 
Together, these frameworks offer a unifying perspective while clarifying the constraints that any classical, quantum, or more general theory must satisfy. 
From an experimental standpoint, this unification raises a central challenge: how can such abstract structures be faithfully realized and tested in realistic physical systems? 
Addressing this question requires translating notions such as measurement compatibility, exclusivity, and global consistency into experimentally accessible protocols, while accounting for imperfections such as finite detection efficiency, noise, and limited measurement sharpness. 
In particular, owing to its high controllability and flexibility, the photonics platform has played a significant role in this field and will be the focus here.

In this Review, we provide a comprehensive overview of the conceptual and experimental connections between quantum contextuality and nonlocality. 
Section~II introduces the sheaf-theoretic framework, emphasizing its unifying logical structure. 
Section~III presents the graph-theoretic approach and its operational implications. 
Section~IV reviews photonic experimental realizations that probe these connections. 
We conclude by outlining open problems and future directions, highlighting how a deeper understanding of the relations between contextuality and nonlocality may inform both fundamental research and the development of quantum technologies. We hope that it can lead to a more systematic approach to harnessing and utilizing both nonlocality and contextuality as resources.

\section{Sheaf-theoretic Approach}

Understanding quantum correlations in a unified and conceptually transparent way requires a framework that treats contextuality and nonlocality on equal footing. These phenomena arise in distinct operational settings: nonlocality involves spatially separated measurements, whereas contextuality involves compatible measurements on a single system. Nevertheless, both originate from the same structural limitation of classical models, which cannot assign globally consistent hidden variables that reproduce the observed correlations.

A powerful unifying framework developed by Abramsky and Brandenburger \cite{abramsky2011sheaf} employs sheaf theory to precisely describe empirical models, in which contextuality and nonlocality are naturally unified and characterized as obstructions to constructing globally consistent descriptions of measurement outcomes. Built on presheaves defined over measurement contexts, this approach reveals that both phenomena arise from a single structural limitation: the impossibility of assigning outcome values that reproduce all locally observed statistics. Cohomology theories can roughly be thought of as descriptions of obstructions to solving some kind of equation. Within this framework, the central distinction lies between local consistency, which requires agreement among overlapping contexts, and global consistency, which demands the existence of a joint assignment for all measurements. The failure to obtain such a global assignment reflects the limitation of classical hidden-variable models, arising from spatial separation in the case of nonlocality and from incompatible measurements in the case of contextuality. 

This section presents a survey of the sheaf-theoretic framework for contextuality and nonlocality. We begin by presenting its basic formal structure and then explain how both phenomena arise as instances of a general obstruction to constructing globally consistent descriptions of measurement statistics. Next we apply the powerful tools of presheaf cohomology to witness and characterise non-locality and contextuality. We conclude by discussing implications for quantum foundations and connections to broader mathematical structures, such as cohomology and logic. An advantage to the sheaf-theoretic framework is that it comes equipped with a particular representation that provides a powerful means of reasoning about empirical models. 

\subsection{The Sheaf-theoretic Framwork}

To study contextuality and nonlocality in a unified way, we need a formalism that captures two essential aspects of quantum experiments: which measurements can be performed together, and how the outcomes are distributed. Operationally, a single run of such an experiment consists of selecting a set of compatible measurements, performing them on a physical system, and observing the resulting outcomes. Each run yields an \emph{event}, defined by the chosen measurements and their outcomes. Repeated trials yield relative frequencies, which for each jointly performable set of measurements define a probability distribution over events. The collection of such distributions across all measurement contexts constitutes an \emph{empirical model}. Formally, one begins by specifying a \emph{measurement scenario} $(X,O,\mathcal{M})$, where $X$ is a finite set of measurements, $O$ is a finite set of outcomes, and $\mathcal{M}$ is a family of \emph{measurement contexts} $C \subseteq X$, representing sets of measurements that can be performed jointly. For each $C \in \mathcal{M}$, specifying a distribution on the events $O^C$. 
To organize a structural description of empirical models, we introduce the mathematical language of \emph{presheaf} and \emph{sheaf}. Let $\mathcal{P}(X)$ denote the poset of subsets of $X$, ordered by inclusion and regarded as a category in the usual way. 
\begin{definition}
\label{presheaf}
A presheaf on $\mathcal{P}(X)$ is a functor 
\begin{equation}
\mathcal{E}: \mathcal{P}(X)^{\mathsf{op}} \to \mathbf{Set}.
\end{equation}
\end{definition}
For each subset $C \subseteq X$, the presheaf $\mathcal{E}$ assigns a set $\mathcal{E}(C)$, and for each inclusion $C' \subseteq C$, a restriction map $\mathcal{E}(C) \to \mathcal{E}(C')$. Functoriality ensures that restrictions compose consistently and that the identity inclusion induces the identity map. Elements of $\mathcal{E}(C)$ are called \emph{sections over $C$}, which represent assignments of outcomes to the measurements in $C$. If $X$ itself is included as an object, a section over $X$ is referred to as a \emph{global section}. 
In the present context, the most important example is the \emph{event presheaf}, defined by $\mathcal{E}(C) := O^{C}$, the set of all outcome assignments to the measurements in $C$. Restriction maps are given by marginalization, corresponding to forgetting outcomes of measurements not contained in the smaller context. A global section of this presheaf therefore corresponds to a deterministic assignment of outcomes to all measurements in $X$. 

The concept of a sheaf refines that of a presheaf by imposing a consistency requirement that local data determine global data uniquely whenever they are mutually compatible. 
\begin{definition}\label{sheaf condition}
A presheaf $\mathcal{E}$ on $\mathcal{P}(X)$ is a \emph{sheaf} if for any family of subsets $\{C_i\}$ with $C = \bigcup_i C_i$, and any family of sections $s_i \in \mathcal{E}(C_i)$ satisfying the compatibility condition: 
\begin{equation}
s_i|_{C_i \cap C_j} = s_j|_{C_i \cap C_j} \quad \forall i,j,
\end{equation}
there exists a unique section $s \in \mathcal{E}(C)$ such that $s|_{C_i} = s_i$ for all $i$.
\end{definition}
A useful intuition is that a presheaf assigns information to a poset such that the assignment on any element can be consistently restricted to lower elements. A sheaf further requires that assignments which exist and are locally compatible on all lower elements can be glued, or lifted, to an assignment on the element itself. Physically, this corresponds to the classical intuition that deterministic outcome assignments for overlapping sets of compatible measurements can always be extended consistently. 

\begin{figure}[h]
\centering
\includegraphics[width=0.4\linewidth]{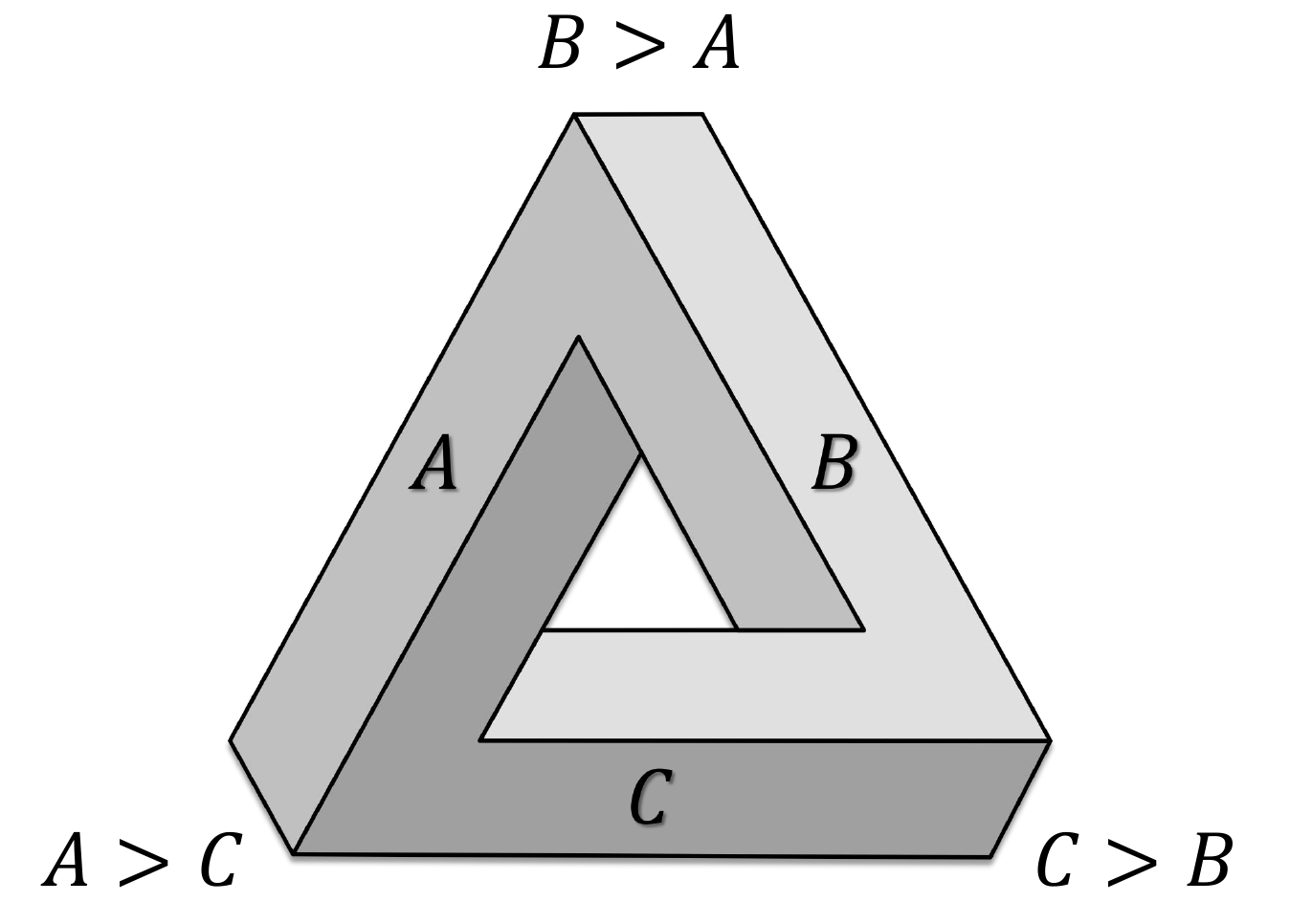}
\caption{Interpreting ``$>$'' as ``appears visually closer,'' the Penrose triangle gives rise to locally consistent orderings that cannot be obtained as the restrictions of any global section \cite{mansfield2013mathematical}. A global section would correspond to a strict total order on $\{A,B,C\}$, which is incompatible with these visual assignments.}
\label{triangle}
\end{figure}

The relevance of this structure to quantum contextuality is immediate. In Kochen-Specker scenarios \cite{kochen1967specker}, it is possible to assign outcomes consistently within each measurement context, yet impossible to construct a single global assignment compatible with all contexts simultaneously. An intuitive geometric analogy is provided by the Penrose triangle in Figure \ref{triangle}, which admits locally consistent constructions but cannot be realized as a single global object. Such situations correspond precisely to families of local sections of the event presheaf that do not arise as restrictions of any global section. This perspective was first articulated in the language of presheaves by Isham and Butterfield~\cite{isham1998topos}, laying the groundwork for the topos approach to quantum theory. While the topos framework incorporates substantial mathematical structure from operator algebras and quantum logic, the sheaf-theoretic approach adopted here is deliberately more operational and minimalist. Its aim is not to reformulate quantum theory itself, but rather to characterize and compare the constraints imposed on empirical data by different physical or hypothetical theories, using only experimentally accessible correlations.
\begin{example}
For a triangle, as show in Figure \ref{triangle}, label the edges by $\{A,B,C\}$, take as poset subsets of the edges labelled by inclusion, and define a presheaf $\mathcal{E}$ that assigns to each subset the set of all strict total orderings of its elements:  
\begin{equation}\label{example1}
\mathcal{E}(\{A, B\})=\{A>B,\, B>A\} .
\end{equation}
Restrictions arise in the obvious way. The triangle represent a family of sections 
\begin{equation}
\left\{s_{\{A, B\}}=B>A,\, s_{\{B, C\}}=C>B,\, s_{\{C, A\}}=A>C\right\},
\end{equation}
which cannot arise from restrictions of any global section $s_{\{A, B\}}$, which in this case would be a strict total order on $\{A, B, C\}$.
\end{example}

To describe empirical models quantitatively, we specify a probability distribution over the assignments $\mathcal{E}(C)$ for each maximal context $C \in \mathcal{M}$. This can be achieved by composing $\mathcal{E}$ with the \emph{distribution functor} $\mathcal{D}_{R} : \mathbf{Set} \to \mathbf{Set}$, which maps a set to the set of $R$-distributions over it, where $R$ is a semiring. Standard probability distributions correspond to $R = \mathbb{R}_{\geq 0}$, the non-negative reals, while other generalizations, such as ‘negative probabilities’ ($R =\mathbb{R}$) \cite{wigner1932quantum,dirac1942bakerian,moyal1949quantum,hiley2012quantum} or ‘possibilistic’ distributions ($R = \mathbb{B}$, the Boolean semiring) \cite{abramsky2011cohomology,fritz2013local}, can also be considered. The composition $\mathcal{D}_{R} \mathcal{E}$ is a presheaf, with restrictions given by marginalization of distributions. 
\begin{definition}
An empirical model $e$ over a measurement scenario $(X,O,\mathcal{M})$ is a family of distributions 
\begin{equation}
\{ e_C \in \mathcal{D}_{R}\mathcal{E}(C) \}_{C \in \mathcal{M}},
\end{equation}
assigning a distribution to each measurement context. 
\end{definition}
Not every such family represents a physically admissible model. Operational consistency requires that distributions associated with different contexts agree on their overlaps:
\begin{equation}
e_C|_{C \cap C'} = e_{C'}|_{C \cap C'} \quad \forall C,C' \in \mathcal{M} .
\end{equation}
This compatibility condition is precisely the sheaf condition restricted to the measurement cover $\mathcal{M}$. In multipartite Bell scenarios it coincides with the no-signaling constraint~\cite{popescu1994quantum}, while in single-system contextuality scenarios it corresponds to the no-disturbance condition~\cite{ramanathan2012generalized}. Empirical models satisfying this condition are therefore referred to as \emph{compatible}. 
The following example illustrates how an empirical model arising from quantum mechanics is represented and analyzed within the sheaf-theoretic framework. 
\begin{example}\label{examplechsh}
Quantum mechanics provides a central class of empirical models, constructed by specifying a quantum state together with a set of observables and evaluating the joint probability distributions of outcomes for each compatible measurement context. As a concrete illustration of this construction, we consider the standard Clauser-Horne-Shimony-Holt (CHSH) scenario \cite{clauser1969proposed}. 
Suppose that two parties, Alice and Bob, share the two-qubit Bell state 
\begin{equation}
\left|\phi^{+}\right\rangle=\frac{1}{\sqrt{2}}\left(|0\rangle_A \otimes|0\rangle_B+|1\rangle_A \otimes|1\rangle_B\right), 
\end{equation}
with each party performs one of two dichotomic measurements. The observables are chosen as 
\begin{equation}
A_1=B_1=\left(\begin{array}{cc}
0 & 1 \\
1 & 0
\end{array}\right), \quad A_2=B_2=\left(\begin{array}{cc}
0 & e^{-i \frac{\pi}{3}} \\
e^{i \frac{\pi}{3}} & 0
\end{array}\right).
\end{equation}
The resulting CHSH empirical model is specified by the joint probability distributions shown in Table~\ref{tab:chsh}. 
\begin{table}[h]
  \centering
   \tabcolsep=4pt
   \renewcommand{\arraystretch}{1.3}
\begin{tabular}{|c|cccc|}
\hline
$P(A, B)$&$A=0, B=0$&$A=0, B=1$&$A=1, B=0$&$A=1, B=1$\\ \hline 
        $A_1,B_1$& $1/2$&$0$& $0$&$1/2$\\ 
        $A_1,B_2$& $3/8$&$1/8$& $1/8$&$3/8$\\ 
        $A_2,B_1$& $3/8$&$1/8$& $1/8$&$3/8$\\ 
        $A_2,B_2$& $1/8$&$3/8$& $3/8$&$1/8$\\ \hline 
\end{tabular}
\caption{The empirical model for the CHSH scenario \cite{clauser1969proposed}.} \label{tab:chsh}
\end{table}

In the sheaf-theoretic formulation, the measurement scenario is given by $X=\left\{A_1,A_2,B_1,B_2\right\}$ and $O=\left\{0,1\right\}$, with the set of maximal contexts 
\begin{equation}
\mathcal{M}=\left\{\{A_1, B_1\}, \left\{A_1, B_2\right\},\left\{A_2, B_1\right\}, \left\{A_2, B_2\right\}\right\}. 
\end{equation}
For each context $C\in\mathcal{M}$, the set $\mathcal{E}(C)$ consists of all outcome assignments $s:C\to O$, 
\begin{equation}
\mathcal{E}(\{A, B\})=\{  A B \mapsto 00, \, A B \mapsto 01,  \,A B \mapsto 10, \, A B \mapsto 11\}.
\end{equation}
Each row of Table~\ref{tab:chsh} specifies a probability distribution over $\mathcal{E}(C)$. For example, the first row corresponds to a distribution
\begin{equation}
e_{\{A_1,B_1\}}\in\mathcal{D}_{R}\mathcal{E}(\{A_1,B_1\}).
\end{equation}
The resulting family of distributions $e=\{e_C\}_{C\in\mathcal{M}}$ constitutes a well-defined sheaf-theoretic empirical model. Since it is derived from quantum mechanics, it satisfies the compatibility (no-signalling) conditions, ensuring that the distributions on overlapping contexts agree on their common marginals.
\end{example}

\subsection{Characterizing Contextuality and Nonlocality}

A central insight of the sheaf-theoretic framework is that both contextuality and nonlocality can be characterized in terms of the \emph{existence of global sections} of a distribution presheaf. In this approach, classicality is identified with the possibility of consistently extending locally observed statistics to a single global probabilistic model defined over all measurements simultaneously. Contextuality and nonlocality arise precisely when such an extension is impossible. 

Let $X$ denote the set of all measurements in a given scenario, and let $\mathcal{E}$ be the event sheaf assigning to each subset $C \subseteq X$ the set $\mathcal{E}(C)=O^C$ of outcome assignments for measurements in $C$. A global section $s\in\mathcal{E}(X)$ specifies a definite outcome for every measurement, independently of the context in which it is performed. Such global assignments therefore correspond to \emph{classical (deterministic) hidden variable models}. 
A probabilistic hidden-variable model is represented by a distribution
$d \in \mathcal{D}_{R}\mathcal{E}(X)$ over global sections. Each deterministic assignment $s\in\mathcal{E}(X)$ induces a Dirac distribution $\delta_s \in \mathcal{D}_{R}\mathcal{E}(X)$, defined by $\delta_s(s)=1$ and $\delta_s(s')=0$ for $s'\neq s$. The restriction of $\delta_s$ to a context $C$ yields the deterministic distribution $\delta_s|_C\in\mathcal{D}_{R}\mathcal{E}(C)$. 
Given a distribution $d$ over global sections, the empirical statistics for each context $C$ are recovered by marginalization, 
\begin{equation}
e_C(s)
  = d|_C(s)
   = \sum_{\substack{s'\in\mathcal{E}(X)\\ s'|_C=s}} d(s') = \sum_{s'\in\mathcal{E}(X)} \delta_{s'|_C}(s)\, d(s'),
\end{equation}
which shows explicitly that the observed probabilities arise from averaging over deterministic hidden variables. For each deterministic assignment $s$ and context $C$, the induced distribution factorizes over individual measurements, 
\begin{equation}
\delta_s|_C(s') = \prod_{x\in C} \delta_{s|\{x\}}\left(s'|\{x\}\right),
\end{equation}
reflecting the fact that deterministic hidden variables assign outcomes to each measurement independently of the context in which it appears. 
These observations lead to the following fundamental equivalence: 
\begin{proposition}
An empirical model admits a global section if and only if it can be realized by a deterministic hidden-variable model.
\end{proposition}

We now reformulate the problem of determining whether an empirical model admits a global section as a linear feasibility question. 
Let $\mathcal{M}$ be a measurement cover of $X$. Consider the disjoint union of local sections, $\coprod_{C \in \mathcal{M}} \mathcal{E}(C)$, and enumerate its elements as $\{s_i\}_{i=1}^{p}$. Similarly, enumerate all global sections of $\mathcal{E}$ as $\{t_j\}_{j=1}^{q}\subseteq\mathcal{E}(X)$. The incidence matrix $\mathbf{M}$ is the $p\times q$ Boolean matrix defined by
\begin{equation}
\mathbf{M}[i, j]= 
\begin{cases}
1, & t_j | C=s_i\left(s_i \in \mathcal{E}(C)\right) \\ 0, & \text {otherwise}
\end{cases}
\end{equation}
Each column of $\mathbf{M}$ corresponds to a global assignment and records the local sections to which it restricts; each row identifies the set of global assignments compatible with a given local section. Equivalently, $\mathbf{M}$ provides a matrix representation of the restriction map 
\begin{equation}
\mathcal{E}(X) \longrightarrow \prod_{C\in\mathcal{M}} \mathcal{E}(C),\quad
s \mapsto (s|_C)_{C\in\mathcal{M}}.
\end{equation}
This construction depends only on the measurement cover $\mathcal{M}$ and the event presheaf $\mathcal{E}$, and is therefore independent of any particular choice of probability theory. It may be applied uniformly to empirical models defined with respect to an arbitrary distribution functor $\mathcal{D}_{R}$. 
Given an empirical model $\{e_C\}_{C \in \mathcal{M}}$ valued in $\mathcal{D}_{R}$, each local section $s_i \in \mathcal{E}(C)$ is assigned a weight $e_C(s_i)$ in the underlying semiring $R$. Collecting these weights over all contexts yields a vector $\mathbf{v} \in R^{p}$, with components $\mathbf{v}[i] = e_C(s_i)$. A putative global distribution is represented by a vector $\mathbf{x} \in R^{q}$, whose entries assign weights to the global sections $\{t_j\}$. 
The requirement that the global distribution reproduces the empirical distributions on every context is encoded by the linear system $\mathbf{M}\mathbf{x} = \mathbf{v}$. To ensure normalization, one may augment $\mathbf{M}$ with an additional row of ones and append a corresponding entry $1$ to $\mathbf{v}$, obtaining an augmented system $\mathbf{M}'\mathbf{x} = \mathbf{v}'$. The existence of a solution $\mathbf{x}$ to this system characterizes whether the empirical model admits a global section. In this way, the question of noncontextuality is recast as a linear feasibility problem, providing a direct and operational criterion for contextuality and nonlocality. 
\begin{proposition}
Solutions of the augmented system $\mathbf{M}'\mathbf{x}=\mathbf{v}'$ in $R$ are in one-to-one correspondence with global sections realizing the empirical model \cite{abramsky2011sheaf}. 
\end{proposition}

\begin{example}

We now analyze the empirical model associated with the CHSH scenario introduced in Example~\ref{examplechsh}. In the sheaf-theoretic framework, the existence of a noncontextual or local hidden-variable model is equivalent to the existence of a probability distribution on the set of global assignments $\mathcal{E}(X)$ whose marginals reproduce the observed empirical data. Concretely, this amounts to finding a solution to a linear system over the reals, subject to the constraints $X_i \ge 0$ and $\sum_i X_i = 1$, where the variables $X_i$ represent the weights assigned to global outcome assignments. 
For the CHSH model \cite{clauser1969proposed}, the marginal constraints implied by the joint probability distributions in Table~\ref{tab:chsh} lead to the following linear equations: 
\begin{equation}
   X_1+X_2+X_3+X_4=1 / 2
\end{equation}
\begin{equation}
   X_2+X_4+X_6+X_8=1 / 8 
\end{equation}
\begin{equation}
   X_3+X_4+X_{11}+X_{12}=1 / 8 
\end{equation}
\begin{equation}  
   X_1+X_5+X_9+X_{13}=1 / 8
\end{equation}
Adding the last three equations yields 
\begin{equation}
X_1+X_2+X_3+2 X_4+X_5+X_6+X_8+X_9+X_{11}+X_{12}+X_{13}=3/8 .
\end{equation}
Since all variables $X_i$ are required to be non-negative, the left-hand side of this equation must be greater than or equal to the left-hand side of the first equation. However, the first equation fixes that sum to be $1/2$, leading to a contradiction. We therefore conclude that no non-negative solution exists. 
This establishes that the CHSH empirical model admits no global probability distribution compatible with all its marginals. In the sheaf-theoretic language, the model has no global section, thereby witnessing Bell nonlocality and the impossibility of a local hidden-variable realization.
\end{example}

We now relate global sections to the more familiar notion of hidden-variable models. Let $\Lambda$ denote a set of hidden variables. A hidden-variable model specifies, for each $\lambda\in\Lambda$ and each context $C\in\mathcal{M}$, a distribution
$h_C^\lambda\in\mathcal{D}\mathcal{E}(C)$, together with a context-independent distribution $h_\Lambda\in\mathcal{D}(\Lambda)$ over hidden variables. For each fixed $\lambda$, the family $\{h_C^\lambda\}_{C\in\mathcal{M}}$ is required to be compatible,
\begin{equation}
h_C^\lambda|_{C\cap C'} = h_{C'}^\lambda|_{C\cap C'} ,
\end{equation}
which corresponds to parameter independence in Bell scenarios. 
We now consider a general notion of hidden-variable models and analyze the precise relationship between classicality, understood as the factorizability of the empirical model, and the existence of global sections. 
A hidden-variable model $h$ is said to realize an empirical model $e$ if the empirical probabilities are recovered by averaging over the hidden variables. Explicitly, for all $C \in \mathcal{M}$ and all $s \in \mathcal{E}(C)$, 
\begin{equation}
e_C(s)  =  \sum_{\lambda \in \Lambda} h^{\lambda}_{C}(s) \cdot h_{\Lambda}(\lambda) .
\end{equation}
The explanatory power of a hidden-variable model lies in its ability to reproduce the statistics of an empirical model, such as the quantum mechanics, while attributing them to underlying variables with more classical behaviour. To achieve this, the model must satisfy the property of \emph{factorizability}, which incorporates both Bell locality and noncontextuality at the level of probability distributions. 
A hidden-variable model $h$ is said to be \emph{factorizable} if, for every context $C \in \mathcal{M}$ and every section $s \in \mathcal{E}(C)$,
\begin{equation}
h^{\lambda}_{C}(s)  = \prod_{m \in C} h^{\lambda}_{C} | \{ m \}  (s | \{ m \}) .
\end{equation}
This condition states that the probability assigned to a joint outcome factors into probabilities assigned to individual measurement outcomes. 
If $m \in C \cap C'$, the compatibility condition implies $h^{\lambda}_{C} | \{ m \}  = h^{\lambda}_{C'} | \{ m \}$, ensuring that the outcome distribution for a single measurement does not depend on the context in which it is measured.
\begin{proposition}
\label{factorizable}
Let $e$ be an empirical model defined on a measurement cover $\mathcal{M}$ for a distribution functor $\mathcal{D}_{R}$. The following are equivalent \cite{abramsky2011sheaf}:
\begin{enumerate}
\item $e$ has a realization by a factorizable hidden-variable model.
\item $e$ has a global section.
\end{enumerate}
\end{proposition}
More intuitively, factorizability means that, conditioned on any value of the hidden variable, the probability assigned to a joint outcome decomposes into a product of probabilities for individual outcomes. Within the sheaf-theoretic framework, this notion admits a precise and unifying formulation. An empirical model $\{e_C\}_{C\in\mathcal{M}}$ is said to be \emph{extendable to a global section} if there exists a distribution $d \in \mathcal{D}_{R}\mathcal{E}(X)$ such that $d|C = e_C$, for all $C \in \mathcal{M}$. The existence of such a global section provides a necessary and sufficient condition for classicality. 
In Bell scenarios, extendability to a global section is equivalent to Bell locality~\cite{bell1964einstein}, while in general measurement scenarios it coincides with noncontextuality in the sense of the Kochen-Specker theorem~\cite{kochen1967specker}. From this viewpoint, nonlocality appears as a special instance of contextuality arising from a particular structure of the measurement cover. The sheaf-theoretic framework therefore yields a single, theory-independent criterion for classical correlations: an empirical model is classical if and only if it admits a global section. Contextuality and nonlocality arise precisely when such a section does not exist. 
This unifying perspective has led to a broad range of structural and conceptual insights \cite{abramsky2011sheaf,abramsky2011cohomology,abramsky2014classification,abramsky2020logic,mansfield2013construction,barbosa2013structural}, and serves as a cornerstone for modern approaches to the study of contextuality and nonlocality.

\subsection{Unified Perspective and Outlook}

The sheaf-theoretic framework provides a unified and explicitly theory-independent characterization of contextuality and nonlocality, formulated entirely in terms of the global consistency structure of empirical models \cite{abramsky2011sheaf,abramsky2011cohomology,abramsky2013relational}. At its core lies a single organizing principle: an empirical model is classically realizable if and only if its locally compatible marginals admit a global section. From this perspective, contextuality and nonlocality arise precisely as obstructions to global extendability, rather than as phenomena tied to any particular mathematical formalism of quantum theory.
This viewpoint recasts the role of hidden-variable theories and measurement incompatibility at a fundamentally structural level. Factorizable hidden-variable models correspond exactly to empirical models that admit global sections, while the failure of such models reflects not the absence of underlying variables, but the impossibility of constructing a jointly consistent assignment compatible with all observable marginals. In this sense, classical explanations fail not because probability theory itself breaks down, but because the local statistical data cannot be glued into a single global probabilistic model. 
Likewise, measurement incompatibility, often attributed to the noncommutativity of quantum observables, admits a more general interpretation. In the sheaf-theoretic framework, it can be understood as the nonexistence of a joint probability distribution whose marginals reproduce the observed empirical statistics~\cite{fine1982hidden,brandenburger2008classification}. In this sense, incompatibility reflects a structural obstruction imposed by global consistency requirements on probability assignments, rather than a specifically quantum postulate. 

Beyond its unifying explanatory role, the sheaf-theoretic framework naturally delineates several promising directions for further development. One important direction is the extension from predominantly finite scenarios to fully measure-theoretic settings, enabling the systematic treatment of continuous-variable and infinite-dimensional systems within the same formalism \cite{giry2006categorical}. A second major challenge concerns multipartite and high-dimensional scenarios, where the combinatorial and algebraic complexity of measurement covers grows rapidly \cite{fritz2013local,navascues2007bounding,sheng2025crypto,sheng2025orbita}. Addressing these regimes may require new categorical, algebraic, or cohomological tools capable of classifying and quantifying obstructions to global consistency in a scalable manner \cite{isham1998topos,doring2008topos,heunen2009topos}. TIn this context, a deeper comparison with topos-theoretic approaches to quantum theory appears particularly promising. While both frameworks interpret contextuality in terms of the failure of global sections, clarifying the precise relationship between operationally defined measurement covers, empirical distributions, and quantum-specific structures such as the spectral presheaf may help distinguish genuinely universal features of quantum correlations from those tied to specific theoretical formalisms. Finally, the framework provides a natural setting for analyzing quasi-probability and negative-probability representations of quantum phenomena \cite{spekkens2008negativity,abramsky2017quantum}. Historically, such representations, ranging from the Wigner function to related phase-space constructions, have signaled the impossibility of realizing a classical joint distribution compatible with all observable marginals, rather than any internal inconsistency of probability theory itself \cite{wigner1932quantum,dirac1942bakerian,moyal1949quantum,hiley2012quantum}. Interpreted sheaf-theoretically, they suggest the possibility of generalized empirical models in which global sections exist only in a quasi-probabilistic sense. Exploring the operational and conceptual significance of such generalized models may further clarify the precise limits of classical explanations of quantum correlations. 

In summary, the sheaf-theoretic framework provides a rigorous and unified characterization of contextuality and nonlocality in terms of global consistency and its obstruction. Far from being a mere technical reformulation of known results, it supplies a flexible and conceptually transparent language for understanding the deep structural relations between contextuality and nonlocality, and offers a promising foundation for future investigations across a wide range of physical theories.

\section{Graph-theoretic Approach}

While the sheaf-theoretic framework provides a high-level and mathematically powerful approach to unify contextuality and nonlocality, it remains largely abstract with respect to the operational constraints, and the connection to experimentally measurable quantities is often nontrivial. In particular, the sheaf-based description does not explicitly represent how individual measurement events exclude one another, nor how these exclusivity relations restrict the strength of attainable correlations. These limitations motivate alternative approaches that retain conceptual rigor while offering greater operational transparency. The graph-theoretic approach provides a complementary perspective by encoding measurement events as vertices and representing pairwise exclusivity by edges, thereby offering a compact and operationally transparent description of the underlying constraint structure. This combinatorial perspective enables a direct comparison between classical, quantum, and more general probabilistic models at the level of the event structure itself. In this section, we provide an overview of this framework systematically, beginning with the construction of exclusivity graphs from operational scenarios and proceeding to the interpretation of the corresponding graph-theoretic constraints.

\subsection{The Graph-theoretic Framwork}

There are two main approaches underlying graph-theoretic formulations of quantum correlations: the compatibility-hypergraph approach and the exclusivity-graph approach. In the compatibility-hypergraph approach, the primary focus is on the compatibility relations among a finite set of measurements, with correlations arising from the structure of jointly measurable contexts; comprehensive accounts of this framework can be found in \cite{acin2015combinatorial,budroni2022kochen,amaral2018graph}. By contrast, the exclusivity-graph approach characterizes correlations in terms of exclusivity relations among a finite set of measurement events. Within this framework, the convex sets of correlations obtained in classical, quantum, and more general probabilistic theories satisfying the \emph{Exclusivity Principle} are naturally related to fundamental convex sets in graph theory. This review will focus on the latter perspective.

To formalize correlations in an experimentally meaningful way, we adopt a general  operational framework . Two types of interventions are assumed to be available: preparations and operations. A fundamental requirement is reproducibility: each preparation and each operation can be implemented arbitrarily many times, allowing outcome probabilities to be estimated from relative frequencies. For a given preparation, an operation may yield several possible outcomes, each occurring with a well-defined probability. 
Correlations are described in terms of probabilities assigned to operationally defined \emph{events}. Two preparations are said to be operationally equivalent if they give rise to identical outcome probabilities for all possible operations; each such equivalence class defines a \emph{state}. The resulting state space is assumed to be convex, with extremal points corresponding to pure states. Operations with multiple outcomes are referred to as \emph{measurements}, and two measurements are equivalent if they generate identical outcome statistics for all states. 
A complete mathematical description of the system therefore consists of a set of states, a set of measurements, and a rule assigning outcome probabilities to each measurement-state pair. Such a description constitutes a \emph{probabilistic model}, and a probability theory is understood as a collection of such models.

We restrict attention to outcome-repeatable measurements, whose outcomes are stable under sequential application to the same physical system. A measurement $j$ is outcome-repeatable if, whenever outcome $k$ is obtained, any subsequent measurement of $j$ yields the same outcome with unit probability. Let $P(k|j)$ denote the probability of outcome $k$ upon performing measurement $j$, with the dependence on the preparation left implicit. In general probabilistic theories, a set of measurements need not admit a joint probability distribution. When such a distribution exists for all states, the measurements are said to be \emph{compatible}. 
An \emph{event} is defined as an equivalence class of outcome-measurement pairs that are assigned the same probability for all states. Thus, outcome $a$ of measurement $x$ and outcomes $b,\dots,c$ of measurements $y,\dots,z$ are equivalent if $P(a|x)=P(b,\dots,c|y,\dots,z)$ for every state. 
Two events are \emph{exclusive} if they correspond to compatible measurements and cannot occur simultaneously. The exclusivity relations of a correlation experiment are encoded in an \emph{exclusivity graph} $\mathcal{G}$, whose vertices represent events and whose edges connect exclusive pairs. This representation encodes the combinatorial structure of events underlying a correlation experiment and provides a unified framework for characterizing classical, quantum, and more general correlations through graph-theoretic quality. 

\begin{example}\label{chsh}

As an illustration of this framework, we reconsider the standard CHSH scenario \cite{clauser1969proposed}, introduced previously in Example~\ref{examplechsh}, with the explicit purpose of constructing its associated exclusivity graph. 
In this scenario, there are four dichotomic measurements labeled $0,1,2,3$, each with outcomes $0$ and $1$. 
Measurements $0$ and $2$ are performed by Alice, while measurements $1$ and $3$ are performed by Bob. 
Joint measurements are performed only for the four compatible measurement pairs $(0,1)$, $(1,2)$, $(2,3)$, and $(3,0)$ on the same prepared states. Each joint outcome $(a,b|x,y)$ constitutes a measurement event and will be represented as a vertex in the graph. 

The exclusivity relation between events is defined operationally: two events are exclusive if they cannot occur simultaneously in a single run of the experiment. In the CHSH scenario, such exclusivity arises whenever two events assign different outcomes to at least one joint measurement. Edges in the graph are introduced to encode this exclusivity relation. Applying these rules yields the full exclusivity graph of the CHSH scenario, denoted by ${\cal G}_{\rm CHSH}$. This graph contains $16$ vertices, corresponding to all joint events associated with the compatible measurement settings, with edges connecting all pairs of mutually exclusive events. 
Maximal sets of pairwise exclusive events form cliques in ${\cal G}_{\rm CHSH}$, that is, each fixed joint measurement setting $(x,y)$ gives rise to a clique of size four, corresponding to the four possible joint outcomes.

In Figure~\ref{graph}, vertices represent events and edges represent exclusivity relations. 
For graphical clarity only, sets of pairwise exclusive events are arranged along straight lines or circumferences. 
The CHSH expression $S_{\rm CHSH}$ involves only a subset of the measurement events.  
Restricting ${\cal G}_{\rm CHSH}$ to the eight events appearing in the inequality, and retaining all exclusivity relations among them, defines an induced subgraph, denoted by $G_{\rm CHSH}$. By construction, such a subgraph is obtained by selecting all the blue vertices with edges between them present in the original graph. The resulting graph $G_{\rm CHSH}$ is isomorphic to the eight-vertex circulant graph $Ci_8(1,4)$. 

\begin{figure}[h]
\centering
\includegraphics[width=0.65\linewidth]{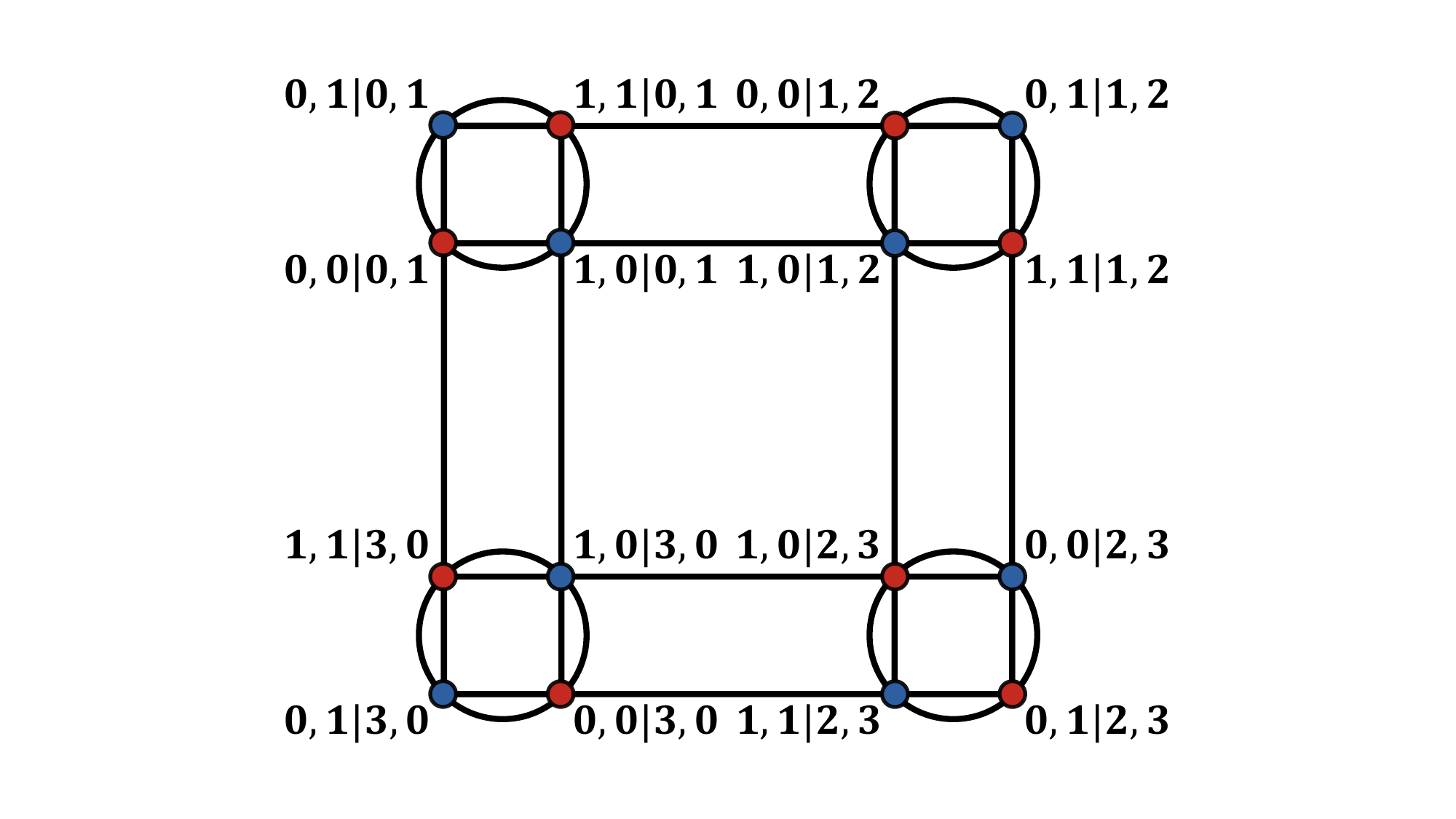}
\caption{The exclusivity graph of the CHSH scenario, ${\cal G}_{\rm CHSH}$ \cite{cabello2014graph}.}
\label{graph}
\end{figure}

The correlations can be expressed as positive linear combinations of event probabilities \cite{badziag2011pentagrams}: 
\begin{equation}
S = \sum_i w_i P(e_i),
\end{equation}
where $w_i > 0$ are weights and $P(e_i)$ denotes the probability of event $e_i$. The corresponding inequalities are: \begin{equation}
 S_{\rm CHSH} = \sum_{i=0}^3 \sum_{a,b} P(a,b \mid i,i+1) \stackrel{\text{LHV}}{\leq} 3,\label{CHSH} 
\end{equation}
the sum over outcomes $a, b \in \{0,1\}$ obeys $a = b$ for $i \neq 2$ and $a \neq b$ for $i = 2$, and the index $i+1$ is taken modulo 4 . The labels ``LHV'' indicate the bounds predicted by local hidden variable models \cite{einstein1935can,bell1964einstein,clauser1969proposed}.

\end{example}

While the example above use uniform weights $w_i = 1$, general expressions may assign different weights to different events. Any such expression $S$ can be encoded as a vertex-weighted exclusivity graph $(G, w)$:
\begin{itemize}
  \item Each vertex $i \in V(G)$ represents an event $e_i$ contributing to $S$;
  \item Edges denote pairs of exclusive events;
  \item The weight function $w: V \to \mathbb{R}_+$ assigns the coefficient $w_i$ to each event.
\end{itemize}
This formalism enables a combinatorial characterization of contextuality and nonlocality, where physical limits on $S$ can be associated with the graph parameters. The exclusivity graph thereby serves as a powerful tool for classifying and analyzing quantum correlations.

\subsection{Combinatorial Bounds on Physical Theories}

Once a correlation experiment is encoded in terms of an exclusivity graph $G$, a natural and fundamental question arises: what ultimately constrains the strength of the correlations that can be assigned to the vertices of $G$? Remarkably, a broad class of physical theories including classical, quantum, and general probabilistic models, can be characterized in terms of the combinatorial structure of $G$, together with a basic probabilistic consistency requirement known as the \emph{exclusivity principle} \cite{cabello2013simple,Yan2013}. 
The exclusivity principle, originally introduced by Specker, asserts that the probabilities associated with any set of pairwise exclusive events must sum to at most 1 \cite{specker2011logik,wright1978mathematical}. In both classical and quantum theories this constraint is automatically satisfied, as it follows from the existence of an underlying probability measure or a quantum state, respectively. However, for general probabilistic models no such guarantee exists, and violations of the exclusivity principle can occur \cite{liang2011specker}. Throughout this section, we denote by $\mathrm{E1}$ the class of theories that satisfy the exclusivity principle when it is imposed only at the level of a single exclusivity graph $G$ \cite{cabello2014graph,Yan2013}.

One of the central results of the graph-theoretic framework is that the maximum value of a correlation expression $S$ attainable within different classes of physical theories is completely characterized by specific graph invariants \cite{cabello2014graph}: 
\begin{equation}
   S \stackrel{\mathrm{LHV,NCHV}}{\leq} \alpha(G,w)
   \stackrel{\mathrm{Q}}{\leq} \vartheta(G,w)
   \stackrel{\mathrm{E1}}{\leq} \alpha^*(G,w).
   \label{bounds}
\end{equation}
Here $\alpha(G,w)$ denotes the independence number of the weighted graph $(G,w)$ \cite{grotschel1986relaxations}, corresponding to classical models admitting deterministic and noncontextual value assignments \cite{kochen1967specker}. The quantity $\vartheta(G,w)$ is the Lov\'asz number \cite{lovasz1979shannon,grotschel1981ellipsoid}, which characterizes the maximum value achievable within quantum theory \cite{budroni2022kochen}. Finally, $\alpha^*(G,w)$ is the fractional packing number \cite{shannon1956zero}, describing the maximal value allowed by theories satisfying the exclusivity principle $\mathrm{E1}$, according to which the probabilities associated with the vertices of any clique of $G$ must sum to at most 1 \cite{cabello2014graph}.

In classical hidden-variable models, both local in Bell scenarios and noncontextual in single-system settings, measurement outcomes are assumed to be predetermined. Consequently, each event is assigned a definite value $0$ or $1$, and probabilities arise as convex mixtures of such deterministic assignments. Since pairwise exclusive events cannot both be assigned value $1$, the set of events simultaneously assigned value $1$ must form an independent (or stable) set of the graph $G$. The maximal classical value of $S$ is therefore given by the maximum total weight of an independent set, which is the weighted independence number $\alpha(G,w)$ \cite{grotschel1986relaxations}. 

In quantum mechanics, exclusivity between events is implemented by the orthogonality of the corresponding projective measurements. To formalize this structure, we adopt the standard graph-theoretic notion of an \emph{orthonormal representation} (OR). 
An orthonormal representation of a graph $G$ in $\mathbb{R}^d$ assigns a nonzero unit vector $\ket{v_i} \in \mathbb{R}^d$ to each vertex $i \in V(G)$ such that $\langle v_i | v_j \rangle = 0$, whenever $i$ and $j$ are nonadjacent in $G$. 
In the framwork of exclusivity graphs, it is convenient to work with the complement $\overline{G}$ of a graph $G$, which is the graph with vertex set $V$ such that two vertices $i,j$ are adjacent in $\overline{G}$ if and only if $i,j$ are not adjacent in $G$. 
Given an orthonormal representation of $\overline{G}$ and a normalized state vector $\ket{\psi}$, referred to as the \emph{handle}, one considers the quantity $\sum_{i \in V(G)} w_i | \langle \psi | v_i \rangle |^2$. The maximum of this expression, taken over all choices of the handle $\ket{\psi}$, is the Lov\'asz number $\vartheta(G,w)$. 
From an operational perspective, this construction captures the quantum realization of events as projection operators $\Pi_i$ acting on a Hilbert space of suitable dimension, with probabilities given by $p_i = \langle \psi | \Pi_i | \psi \rangle$. For rank-one projectors, one may identify $\ket{v_i} = \Pi_i \ket{\psi}/\sqrt{\langle \psi | \Pi_i | \psi \rangle}$, which yields an orthonormal representation of $\overline{G}$ with handle $\ket{\psi}$. Consequently, the maximal value of the correlation expression $S$ attainable within quantum theory coincides with the Lov\'asz number $\vartheta(G,w)$ \cite{lovasz1979shannon}. 

For general probabilistic theories constrained only by the exclusivity principle $\mathrm{E1}$, the maximal value of the correlation expression $S$ is given by the fractional packing number $\alpha^*(G,w) := \max \sum_{i \in V(G)} w_i\, p_i$, where the maximization is taken over all assignments $p_i \ge 0$ satisfying the clique constraints $\sum_{i \in C} p_i \le 1$ for every clique $C \subseteq G$. 
Each clique corresponds to a set of pairwise exclusive events, and these linear constraints implement the exclusivity principle directly at the probabilistic level. The quantity $\alpha^*(G,w)$ therefore characterizes the strongest correlations compatible with the exclusivity structure encoded by $G$ under this minimal consistency requirement.

The three graph parameters appearing in Equation~\eqref{bounds} differ markedly in their computational complexity. Determining the independence number $\alpha(G,w)$ is NP-hard in general, even in the unweighted case \cite{grotschel2012geometric}. By contrast, the Lov\'asz number $\vartheta(G,w)$ can be computed to arbitrary precision in polynomial time using semidefinite programming \cite{grotschel1986relaxations}. 
The fractional packing number $\alpha^*(G,w)$ is defined by a linear program with clique constraints. While the optimization problem itself is efficiently solvable once the relevant clique inequalities are specified, generating the full list of cliques of an arbitrary graph is NP-hard in general \cite{grotschel1986relaxations}.

\begin{example}
The unifying power of the graph-theoretic framework becomes transparent when applied to paradigmatic correlation inequalities. In the CHSH scenario \cite{clauser1969proposed}, the associated exclusivity graph yields the bounds $\alpha = 3$, $\vartheta = 2 + \sqrt{2}$, and $\alpha^* = 4$, corresponding respectively to the maximal values attainable within local hidden-variable theories \cite{clauser1969proposed}, quantum theory \cite{cirel1980quantum}, and no-signaling models \cite{popescu1994quantum}. 
In the Klyachko-Can-Binicio\u{g}lu-Shumovsky (KCBS) scenario \cite{Klyachko2008}, one similarly finds $\alpha = 2$, $\vartheta = \sqrt{5}$, and $\alpha^* = 5/2$, reproducing the established bounds for noncontextual hidden-variable theories \cite{Klyachko2008}, quantum theory \cite{badziag2011pentagrams}, and generalized no-disturbance models \cite{ramanathan2012generalized}. Despite their distinct operational and physical interpretations, both scenarios are captured by the same combinatorial hierarchy encoded in Equation~\eqref{bounds}.
\end{example}

The Lov\'asz number $\vartheta(G,w)$ provides a universal upper bound on the quantum value of any Bell or noncontextual inequality associated with a vertex-weighted exclusivity graph $(G,w)$. A fundamental result of the graph-theoretic approach shows that, for every graph $G$, there always exists a noncontextuality inequality whose maximum quantum violation is exactly given by $\vartheta(G)$. Moreover, in such scenarios, the set of quantum-achievable probability assignments to the vertices of $G$ coincides with the Gr\"otschel-Lov\'asz-Schrijver theta body ${\rm TH}(G)$ \cite{grotschel1986relaxations}. 
However, this correspondence does not always hold in Bell scenarios. Not every graph $G$ admits a Bell inequality that achieves the quantum bound $\vartheta(G)$, due to additional constraints imposed by Bell scenarios, which enforce a specific labeling of measurement events \cite{sadiq2013bell}. This provides further evidence that nonlocality constitutes a special subclass of contextuality. To some extent, the notion of contextuality abstracts away from specific experimental realizations and thereby significantly simplifies the characterization of the quantum set, compared with approaches that rely on a fixed labeling of measurement events \cite{cirel1980quantum,navascues2007bounding,navascues2008convergent}.

In analogy with the definition of the quantum set ${\cal Q}(G)$, one can introduce the corresponding sets for classical theories and for more general probabilistic theories satisfying the exclusivity principle $\mathrm{E1}$. 
We denote by ${\cal C}(G)$ the convex hull of all deterministic noncontextual assignments, 
\begin{equation}
{\cal C}(G) := \mathrm{conv}\left\{ x^S : S \text{ is a stable set of } G \right\},
\end{equation}
where a deterministic assignment $x^{S}\in\{0,1\}^{|V|}$ is associated with a stable set $S\subseteq V(G)$ via $x^{S}_i=1$ if $i\in S$ and $x^{S}_i=0$ otherwise. Similarly, we define ${\cal E}^{1}(G)$ as the polytope of all probability assignments satisfying the exclusivity principle $E1$, 
\begin{equation}
{\cal E}^1(G) := \left\{ p \in \mathbb{R}_+^{|V|} : \sum_{i \in C} p_i \le 1 \text{ for all cliques } C \right\},
\end{equation}
With these definitions, one obtains the inclusion 
\begin{equation}
{\cal C}(G) \subseteq {\cal Q}(G) \subseteq {\cal E}^1(G).
\end{equation}
Both ${\cal C}(G)$ and ${\cal E}^{1}(G)$ are polytopes, whereas ${\cal Q}(G)$ is generally not. 
A natural question is under which conditions distinguish quantum from classical correlations, and under which conditions the exclusivity principle $\mathrm{E1}$ singles out the quantum set. It is proven that ${\cal C}(G) = {\cal Q}(G)$ and ${\cal Q}(G) = {\cal E}^{1}(G)$ hold if and only if the graph $G$ contains no induced odd cycle $C_n$ with $n\ge 5$, nor the complement $\overline{C_n}$ of such a cycle \cite{cabello2014graph,grotschel1986relaxations,berge1961farbung}. Remarkably, the answer is entirely graph-theoretic and is characterized by the absence of certain induced subgraphs. 

\subsection{Summary and Further Perspectives}

The graph-theoretic approach provides a conceptually transparent and operationally grounded framework for analyzing quantum correlations. By taking measurement events as fundamental and encoding their mutual exclusivity in a graph, one abstracts away from specific experimental setups while retaining the essential constraints that any probabilistic theory must satisfy. In this sense, the exclusivity graph is not merely a mathematical convenience, but the natural language for expressing the structure of correlations and the limits imposed by classical, quantum, or more general probabilistic theories. 

A central insight of this framework is that the maximal correlations attainable under different theories are determined by intrinsic combinatorial properties of the graph. 
Classical correlations are bounded by the independence structure of the graph, reflecting the limitations of predetermined, noncontextual assignments. Quantum correlations surpass classical bounds, with orthogonality imposing geometric consistency, the Lov\'asz number provides as a precise quantification of this quantum advantage. More general probabilistic theories are constrained only by the exclusivity principle, which enforces logical consistency among mutually exclusive events. In this way, the framework connects operational, geometric, and combinatorial features. 

Beyond these formal results, the graph-theoretic perspective highlights the conceptual unity of contextuality and nonlocality. Both phenomena arise from the impossibility of assigning globally consistent truth values to all events, with quantum mechanics allowing correlations that surpass classical bounds while remaining limited by the exclusivity structure encoded in the graph. This perspective unifies previously distinct notions, emphasizing that the distinctive features of quantum correlations are fundamentally about the combinatorial structure of mutually exclusive events. 

Looking forward, the graph-theoretic approach highlights several open questions and promising directions. In multipartite scenarios, additional causal, spatial, or labeling constraints may further restrict the set of achievable quantum correlations, emphasizing the connections between the graph-theoretic principle and experimental structure. More broadly, one may ask whether principles based solely on exclusivity, perhaps supplemented by consistency requirements under graph composition, are sufficient to uniquely characterize the quantum set within the broader landscape of no-disturbance or no-signaling models. Thus, the graph-theoretic framework not only provides a powerful tool for designing and classifying correlation experiments, but also offers a unified perspective on the logical, physical, and structural essence of quantum contextuality and nonlocality. 

\section{Experimental Aspects}

The conceptual unification of contextuality and nonlocality, as formalized within the sheaf-theoretic and graph-theoretic framework, provides a powerful abstraction for characterizing nonclassical correlations independently of Hilbert space representations. While this theoretical framework captures the logical and topological structure of quantum correlations, experimental realization is crucial for validating these insights and for exploring their practical consequences. The experimental challenge lies in translating the abstract notions of global sections, measurement compatibility, and exclusivity into physically implementable protocols, often constrained by measurement sharpness, detection efficiency, and noise. In the remainder of this section, we review several important recent photonics experiments that investigate the connections between contextuality and nonlocality. 

\subsection{Revealing Nonlocality from Local Contextuality}

A particularly instructive class of experiments demonstrates how Bell nonlocality can be operationally revealed from single-system contextuality through sequential compatible measurements. Following a proposal by Cabello~\cite{cabello2010proposal}, Liu \textit{et al.}~\cite{liu2016nonlocality} showed that Bell nonlocality can be revealed by combining locally contextual correlations with bipartite correlations that, taken alone, remain compatible with local hidden-variable (LHV) models. n this sense, nonlocality is not directly exhibited by standard Bell inequality violations, but is instead activated through the presence of state-independent contextuality in one of the subsystems. 

\begin{figure}[h]
\centering
\includegraphics[width=0.6\linewidth]{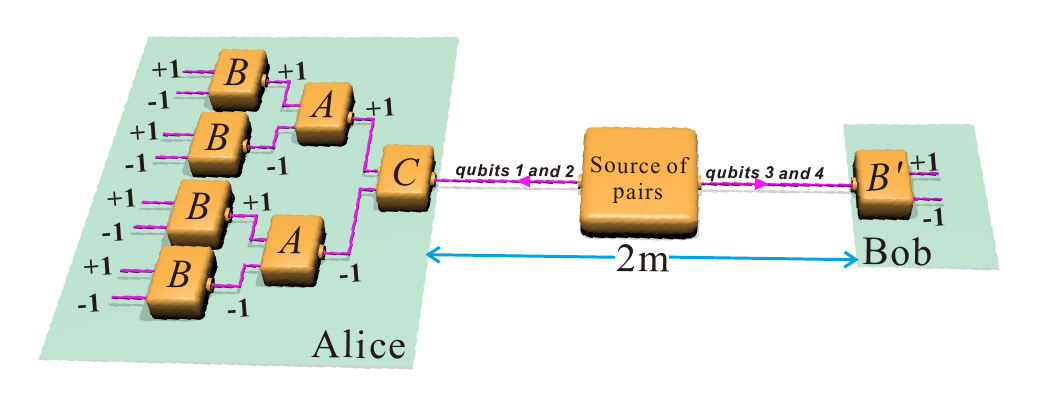}
\caption{\sf Scheme of the experiment by Liu \textit{et al.}~\cite{liu2016nonlocality}. A source emits pairs of entangled particles. One particle encodes two qubits (qubits 1 and 2) and is sent to Alice's laboratory, where three sequential measurements are performed $C$, $A$, and $B$). The other particle, encoding qubits 3 and 4, is sent to Bob's laboratory, where a single measurement is performed ($B'$). }
\label{liu2016}
\end{figure}

The experiment involves four qubits distributed between two distant parties. Alice holds qubits $1$ and $2$, while Bob holds qubits $3$ and $4$. In each experimental run, Alice performs a sequence of three jointly compatible measurements on her subsystem, whereas Bob performs a single measurement on his subsystem, as illustrated in Figure \ref{liu2016} (a). Space-like separation between Alice’s and Bob’s laboratories ensures that no communication can influence their respective measurement choices or outcomes. 
Experimentally, this configuration is realized using two hyperentangled photons, with spatial and polarization degrees of freedom encoding independent qubits.

Alice’s sequential measurements are chosen from the Peres-Mermin set \cite{peres1990incompatible,mermin1990simple}, which exhibits state-independent contextuality. Six sequences of three compatible observables generate local triple correlations, while bipartite correlations arise from comparing Alice’s later measurements in each sequence with Bob’s single measurement. 
The central quantity tested experimentally is the inequality introduced in~\cite{cabello2010proposal}, 
\begin{equation}\label{cabello2010}
\langle\omega\rangle \equiv\langle\chi\rangle+\langle S\rangle \stackrel{\mathrm{LHV}}{\leq} 16,
\end{equation}
which must be satisfied by all local hidden variables (LHV) theories \cite{einstein1935can,bell1964einstein}. The term $\langle \chi \rangle$ consists exclusively of correlations among Alice’s sequential measurements and corresponds to a Peres-Mermin contextuality inequality $\left\langle \chi \right\rangle \leq 4$ \cite{peres1990incompatible,mermin1990simple}, while $\langle S \rangle$ contains only bipartite correlations between Alice’s later measurements in each sequence and Bob’s single measurement \cite{cabello2010proposal}. Experimentally, a strong violation of the noncontextual bound is observed in the local term $\left\langle \chi \right\rangle $, while the bipartite correlations $\langle S \rangle$ alone remain compatible with LHV models. Only when both contributions are jointly considered does the inequality $\langle \omega \rangle \leq 16$ become violated, thereby revealing Bell nonlocality \cite{liu2016nonlocality}. This feature distinguishes the present scheme from standard Bell tests. 

From a conceptual perspective, this experiment does not introduce a new form of nonlocality. Rather, they provide an explicit operational mechanism by which Bell-type violations can emerge from locally contextual measurement structures when distributed across subsystems via entanglement. Although such implementations rely on assumptions of compatibility and non-disturbance inherent to sequential measurements, they establish a clear and experimentally accessible bridge between quantum contextuality and Bell nonlocality. 
From a sheaf-theoretic perspective, Alice’s sequential compatible measurements define a contextual empirical model that admits no consistent global assignment of outcomes, independently of the quantum state. The entangled distant system does not introduce additional contextual structure, but rather redistributes this incompatibility into a bipartite measurement scenario, where it becomes accessible to experimental tests of Bell-type constraints. 
In graph-theoretic terms, the Peres-Mermin measurements implemented by Alice correspond to an exclusivity structure whose quantum correlations exceed the limits imposed by noncontextual models. The Cabello's inequality \eqref{cabello2010} can then be understood as a constraint on a composite exclusivity structure, obtained by embedding this local contextual structure into a larger bipartite scenario through perfect Alice-Bob correlations. The observed violation reflects the impossibility of a single classical probability assignment over the combined set of measurement events.
Within the unified sheaf- and graph-theoretic frameworks discussed above, these experiments exemplify how contextuality can be regarded as a more primitive form of nonclassicality, whose structural incompatibility with classical probability theory may be redistributed across subsystems and ultimately manifested as Bell nonlocality under suitable operational conditions.

\subsection{Coexistence of Contextuality and Nonlocality}

The possibility that nonlocality and contextuality might manifest within a single physical scenario was first systematically analyzed by Kurzy\'nski \textit{et al.}~\cite{kurzynski2014fundamental}. In their proposal, two spatially separated parties perform a standard CHSH Bell test \cite{clauser1969proposed}, while one party simultaneously probes state-dependent contextuality through a KCBS inequality \cite{Klyachko2008} on a local subsystem, using the same set of compatible measurements. Within quantum theory, as well as in more general no-disturbance theories, this scenario was shown to exhibit a trade-off relation between the corresponding nonlocality and contextuality witnesses, implying that at most one of the two inequalities can be violated \cite{kurzynski2014fundamental}. This result motivated the conjecture of a fundamental \emph{monogamy} relation between Bell nonlocality and state-dependent contextuality. 
The conjectured monogamy was subsequently confirmed experimentally~\cite{zhan2016realization}. However, that such constraints do not apply to state-independent contextuality, whose violation does not rely on a specific state preparation. Indeed, simultaneous violations of Bell inequalities and state-independent Kochen-Specker inequalities were later reported~\cite{hu2018simultaneous}. Moreover, the original monogamy argument implicitly relied on the standard Bell scenario in which each party performs a single measurement per experimental run. As pointed out in~\cite{temistocles2019measurement}, each measurement appearing in a CHSH inequality can be in principle supplemented with an additional compatible observable, leading to the formulation of more general Bell inequalities incorporating measurement contexts. 

\begin{figure}[h]
\centering
\includegraphics[width=0.8\linewidth]{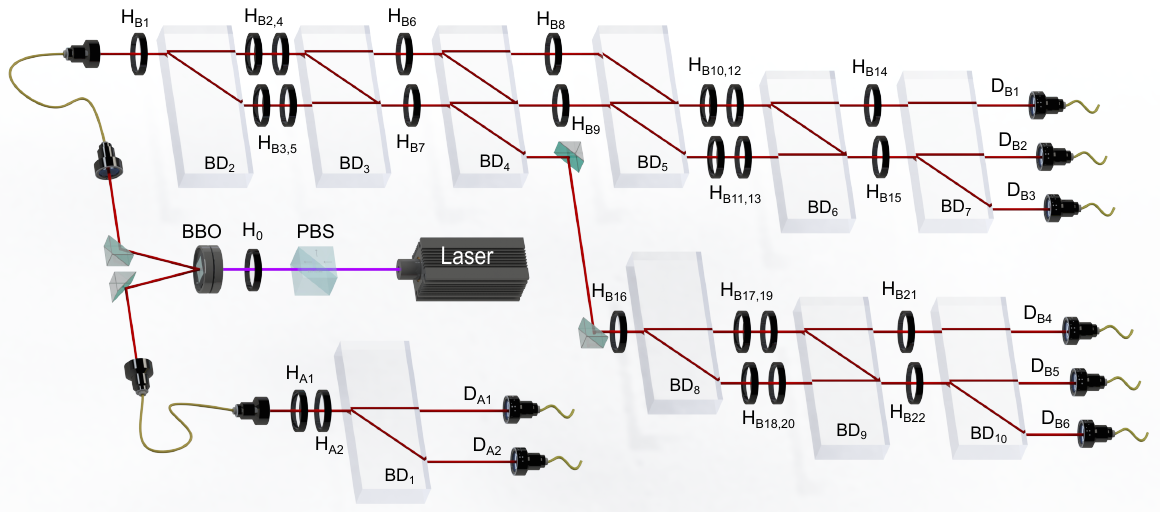}
\caption{\sf Scheme of the experiment by Xue \textit{et al.}~\cite{xue2023synchronous}, illustrating the synchronous test of Bell nonlocality and state-dependent contextuality. Polarization-entangled photon pairs are generated via spontaneous parametric down-conversion. One photon of each pair encodes a qubit in its polarization degree of freedom and is measured by Alice using standard polarization analysis, implementing the CHSH test with Bob. The other photon encodes a qutrit in its combined polarization and spatial degrees of freedom and is measured by Bob through a sequence of compatible measurements realized with cascaded Mach-Zehnder interferometers, enabling the test of the KCBS inequality. The experimental architecture allows both inequalities to be evaluated within the same operational scenario.}
\label{xue}
\end{figure}

Building on these insights, recent experiments have demonstrated that Bell nonlocality and contextuality can in fact coexist within a single operational setting. A representative example is the photonic experiment reported by Xue \textit{et al.}~\cite{xue2023synchronous}, in which the measurement architecture allows for the synchronous evaluation of both the CHSH and KCBS inequalities \cite{clauser1969proposed,Klyachko2008}. By exploiting a generalized Bell scenario with compatible measurements on one subsystem, the experiment achieves simultaneous violations of both inequalities within the same experimental run, in quantitative agreement with quantum mechanical predictions. These results provide a clear proof-of-principle that Bell nonlocality and state-dependent contextuality need not be mutually exclusive, once measurement contexts are treated as integral components of the nonlocality test. 

From a broader perspective, coexistence experiments highlight that the apparent tension between contextuality and nonlocality is not a fundamental limitation, but rather a consequence of restrictive assumptions about measurement structure. When viewed through the unified sheaf- and graph-theoretic frameworks discussed above, these experiments illustrate that contextuality and nonlocality correspond to distinct manifestations of the same underlying obstruction to classical probabilistic models, which may emerge concurrently when the operational scenario admits both incompatible measurements and spatial separation.

\subsection{Converting Contextuality into Nonlocality}

A critical step toward a fully operational unification of contextuality and nonlocality was proposed by Cabello~\cite{cabello2021converting}, who showed that any state-independent contextuality (SI-C) set can be systematically embedded into a bipartite Bell scenario. In this construction, one party measures the projectors defining the SI-C set, while the other party measures their complex conjugates. The resulting Bell inequalities are violated by quantum correlations whenever the underlying measurement set is state-independent contextual, thereby allowing contextuality to be manifested through spacelike separated measurements. Conceptually, this approach converts a single-system logical obstruction into a bipartite manifestation of correlations, while bypassing the long-standing ``compatibility'' and ``sharpness'' loopholes \cite{szangolies2013tests,szangolies2015testing,vallee2024corrected,spekkens2014status,kunjwal2015kochen,krishna2017deriving} inherent to sequential-measurement tests of contextuality \cite{kirchmair2009state,guhne2010compatibility}. 

Despite its generality, the experimental realization of Cabello’s conversion scheme is demanding. The test of SI-C sets require well-controlled projective measurements in Hilbert spaces of corresponding dimension, high-purity entangled states, and precise suppression of noise and mode crosstalk. 
These requirements have long hindered experimental implementations of verifing high-dimensional contextuality. High-dimensional photonic orbital angular momentum (OAM) platforms provide a natural solution to these challenges. OAM modes can be used to define an infinitely dimensional discrete Hilbert space, and the number of effective dimensions can be readily tailored as required \cite{mair2001entanglement,torres2003quantum}, and spontaneous parametric down-conversion naturally produces photon pairs entangled in OAM with correlations \cite{leach2010quantum}. Moreover, spatially separated measurements allow the required projective measurements to be performed without invoking sequential compatibility assumptions \cite{sheng2025orbital}.

Building on these capabilities, Sheng \textit{et al.}~\cite{sheng2025orbital} have experimentally violated Bell inequalities derived from minimal state-independent contextuality (SI-C) sets using high-dimensional OAM-entangled photon pairs. These include the simplest SI-C set in dimension $d=3$ identified by Yu and Oh~\cite{Yu2011}, the minimal Kochen-Specker (KS) set in dimension $d=4$ proposed by Cabello \textit{et al.}~\cite{cabello1996bell}, and the KS set with the smallest number of contexts in dimension $d=6$ introduced by Lison{\v{e}}k \textit{et al.}~\cite{lisonvek2014kochen}. By combining entanglement concentration techniques with optimized holographic measurement strategies, these experiments achieved statistically significant violations of the corresponding Bell inequalities across multiple dimensions. Importantly, the observed violations cannot be attributed to bipartite nonlocal correlations alone, but directly reflect the contextual structure encoded in the underlying SI-C sets.

\begin{figure}[h]
\centering
\includegraphics[width=0.8\linewidth]{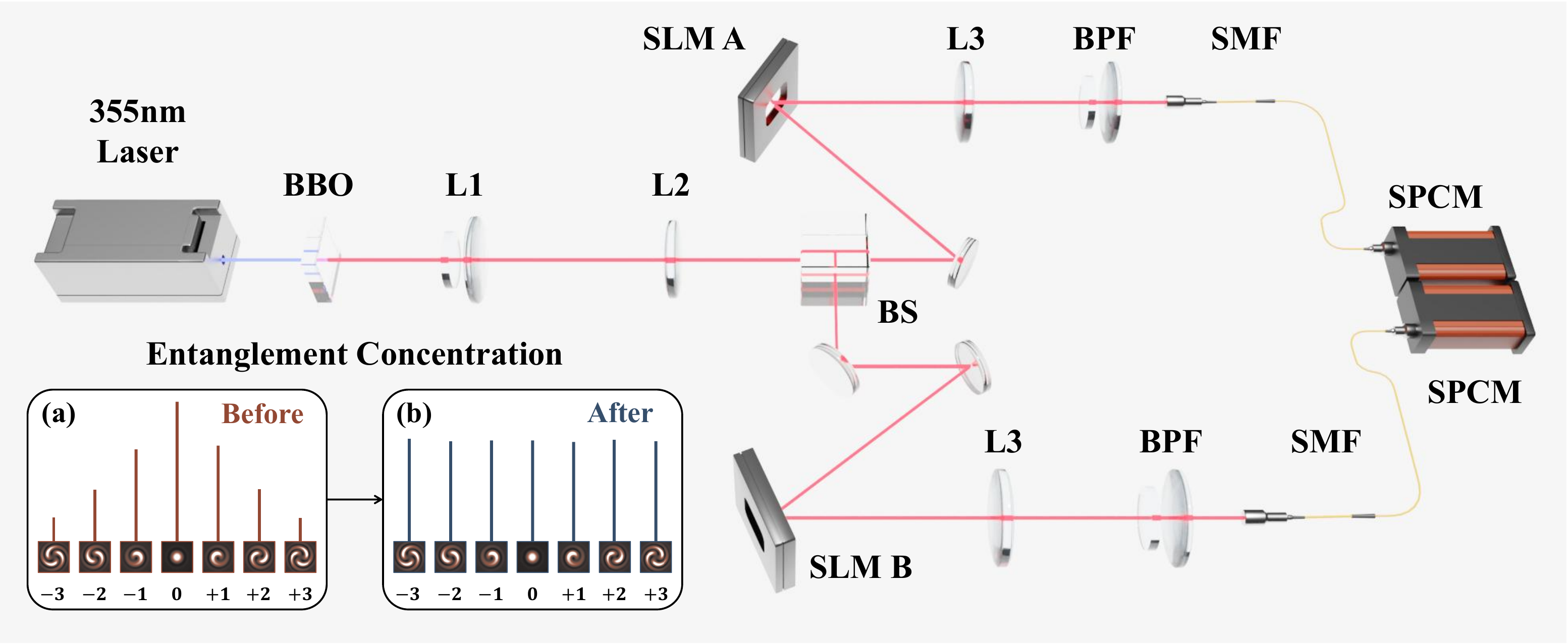}
\caption{\sf Scheme of the experiment by Sheng \textit{et al.}~\cite{sheng2025orbital}. The inset (a) shows the original two-photon OAM spectrum of limited spiral bandwidth before entanglement concentration, while (b) shows the maximally entangled OAM spectrum after concentration.}.
\label{sheng}
\end{figure}

From a unified theoretical perspective, Cabello ’s conversion scheme admits a transparent interpretation within both the sheaf-theoretic and graph-theoretic frameworks discussed above. In sheaf-theoretic terms, a SI-C set defines an empirical model that admits no global section independently of the quantum state. The bipartite embedding redistributes this obstruction across spatially separated measurements, rendering it testable as a Bell-type constraint. In graph-theoretic language, the orthogonality relations of the SI-C set define an exclusivity graph whose noncontextual bound is given by its independence number; the associated Bell inequality inherits this bound, while quantum correlations attain values governed by the corresponding Lov\'asz number. Experimental implementations then realize these abstract structures by engineering entangled states and measurement architectures that faithfully reproduce the required exclusivity relations. 
The experiments demonstrate that contextuality can function as a more primitive quantum resource from which nonlocality can be operationally derived. Unlike earlier approaches that reveal nonlocality through auxiliary contextual correlations or demonstrate their coexistence under generalized measurement scenarios, Cabello’s framework provides a direct and loophole-resistant route for certifying contextuality via genuine Bell tests. In this sense, high-dimensional photonic platforms, particularly those based on OAM entanglement, emerge as a conceptually clean and experimentally powerful arena for realizing the deepest structural connections between contextuality and nonlocality. 

Taken together, the experiments reviewed in this section demonstrate how the abstract unification of contextuality and nonlocality articulated within the sheaf-theoretic and graph-theoretic frameworks can be progressively translated into concrete laboratory protocols. The sheaf-based perspective identifies contextuality and nonlocality as manifestations of a common obstruction to global assignments, while graph-theoretic formulations encode these obstructions into experimentally testable inequalities defined by exclusivity relations. Experimental implementations then realize these abstract structures by engineering specific measurement architectures and quantum states that faithfully reproduce the underlying compatibility and exclusivity patterns. Early experiments established operational bridges in which locally contextual structures give rise to Bell-type violations under suitable correlation analyses, while subsequent coexistence experiments demonstrated that contextuality and nonlocality can be simultaneously observed within a single composite system. More recently, conversion-based schemes grounded in state-independent contextuality have shown that contextual correlations can be certified through genuinely bipartite Bell tests, thereby bypassing key loopholes associated with sequential measurements. In this way, experimental progress has followed a clear conceptual trajectory: from logical and topological characterizations, through combinatorial witnesses, to operational protocols that render the unified structure of quantum correlations empirically accessible.

\section{Conclusions}

In this review, we have developed a unified perspective on quantum contextuality and nonlocality by tracing a coherent progression from abstract structural characterizations to operational formulations and experimental realizations. The sheaf-theoretic framework provides a rigorous, representation-independent characterization of contextuality and nonlocality, highlighting the logical and topological obstructions to classical explanations through the language of global sections and presheaves. This high-level perspective reveals the intrinsic incompatibilities of quantum events, offering a unifying conceptual lens for understanding nonclassical correlations.  

Building on this foundation, the graph-theoretic framework translates the abstract structural insights into combinatorial and operationally meaningful terms. By representing measurement events as vertices and their exclusivity relations as edges, this framework captures classical, quantum, and generalized probabilistic theories bounds through graph invariants such as the independence number, the Lov\'asz number and the fractional packing number. In doing so, it bridges formal theory and experimental practice, providing systematic prescriptions for constructing noncontextuality and Bell inequalities, as well as quantitative predictions for their quantum violations. 

Sheaf-theoretic and graph-theoretic approaches offer complementary perspectives on contextuality and nonlocality. The sheaf framework encodes measurement scenarios as presheaves, with local sections representing compatible event assignments and the existence of a global section signaling noncontextuality. Graph-theoretic methods, in contrast, map events to vertices and exclusivity relations to edges, with independent sets and graph invariants, quantifying classical, quantum, and generalized bounds. These two formalisms are closely connected: the set of local sections in the sheaf approach corresponds to independent sets in the graph, and the Lov\'asz theta number provides a concrete numerical reflection of the cohomological obstructions that indicate contextuality in the sheaf framework. While sheaf theory excels in generality and formal rigor, the graph approach translates abstract structural constraints into operationally accessible and computationally tractable quantities, facilitating experimental design and numerical analysis. Together, they provide a unified understanding, with sheaf theory clarifying the underlying conceptual structure and graph theory enabling practical quantification. 

The sheaf-theoretic and graph-theoretic approaches thus offer complementary yet tightly connected perspectives. In the sheaf framework, contextuality is characterized by the incompatibility of local sections with any global assignment, whereas in the graph-theoretic formulation, classical correlations correspond to independent sets and their convex hulls, with the independent number provides a concrete numerical manifestation of these structural obstructions, translating abstract consistency failures into experimentally testable bounds. While sheaf theory excels in generality and formal rigor, the graph approach translates abstract structural constraints into testable quantities, facilitating experimental design and numerical analysis. Together, they provide a coherent framework in which conceptual structure and quantitative analysis reinforce one another. 

Recent experimental advances, with a particular focus on photonic implementations, have demonstrated the practical relevance of this unified viewpoint, from sequential measurements on single photons to high-dimensional entanglement experiments. These implementations validate the operational content of both sheaf- and graph-based descriptions and, crucially, reveal explicit pathways for relating contextuality and nonlocality. In particular, the embedding of state-independent contextuality sets into bipartite Bell scenarios exemplifies how abstract unification directly informs experimental design, enabling robust and scalable tests of high-dimensional quantum correlations. 

In summary, the progression from sheaf-theoretic characterization to graph-theoretic quantification and experimental realization unifies previously distinct manifestations of quantum nonclassicality within a single structural narrative. Beyond deepening our conceptual understanding, this integrated framework provides guidance for future investigations into higher-dimensional and multipartite systems, device-independent scenarios, and quantitative trade-offs between different forms of quantum correlations. Clarifying whether a fully unified, theory-independent resource description of contextuality and nonlocality can be achieved remains an open challenge, one that promises to further illuminate the foundational structure of quantum theory and its applications in quantum information science.

\medskip


\medskip

%

\bibliographystyle{MSP}




\end{document}